\documentclass[journal]{IEEEtran}
\usepackage{cite}
\usepackage{mciteplus}
\usepackage{citesort}
\usepackage{amsthm}
\theoremstyle{plain} 

\usepackage{graphicx}

\usepackage{epsfig}
\usepackage{epstopdf}
\usepackage{psfrag}
\usepackage{subfigure}

\usepackage{url}

\usepackage{stfloats}
\usepackage{amsmath}
\usepackage{amsfonts}
\usepackage{amssymb}
\usepackage{eqnarray}
\usepackage{subeqnarray}
\usepackage{flushend}

\usepackage{array}

\usepackage{setspace}
\usepackage{paralist}

\usepackage{acronym}

\acrodef{sm}[SM]{spatial modulation}
\acrodef{ssk}[SSK]{space shift keying}
\acrodef{smx}[SMX]{spatial multiplexing}
\acrodef{ml}[ML]{maximum--likelihood}
\acrodef{sd}[SD]{Sphere Decoder}
\acrodef{aber}[ABER]{average bit error ratio}
\acrodef{smrx}[SM--Rx]{Receiver--centric SD}
\acrodef{smtx}[SM--Tx]{Transmit--centric SD}
\acrodef{mimo}[MIMO]{multiple--input multiple--output}
\acrodef{simo}[SIMO]{single--input multiple--output}
\acrodef{cdf}[CDF]{cumulative distribution function}
\acrodef{bpsk}[BPSK]{binary phase shift keying}
\acrodef{qpsk}[QPSK]{quadrature phase shift keying}
\acrodef{qam}[QAM]{quadrature amplitude modulation}
\acrodef{snr}[SNR]{signal-to-noise-ratio}
\acrodef{iid}[i.i.d.]{identical and independently distributed}
\acrodef{ici}[ICI]{inter--channel interference}
\acrodef{rf}[RF]{radio frequency}
\acrodef{nlos}[NLOS]{Non Line--of--Sight}
\acrodef{los}[LOS]{line of sight}
\acrodef{scr}[SC]{spatial correlation}
\acrodef{pep}[PEP]{pairwise error probability}
\acrodef{mgf}[MGF]{moment-generation function}
\acrodef{fbesm}[FBE--SM]{fractional bit encoding spatial modulation}
\acrodef{gsm}[GSM]{generalized spatial modulation}
\acrodef{csi}[CSI]{channel state information}
\acrodef{pdf}[PDF]{probability distribution function}
\acrodef{mrc}[MRC]{maximum ratio combining}
\acrodef{rv}[RV]{random variable}
\acrodef{awgn}[AWGN]{additive white Gaussian noise}
\acrodef{qsm}[QSM]{quadrature spatial modulation} 
\acrodef{mac}[MAC]{multiple access channel}
\acrodef{zmcgv}[ZMCGV]{zero--mean complex Gaussian vector}
\acrodef{dof}[DoF]{degree of freedom}
\acrodef{eda}[OSA]{optimally spaced antennas}
\acrodef{rda}[RSA]{randomly spaced antennas}
\acrodef{aoa}[AOA]{angle of arrival}
\acrodef{aod}[AOD]{angle of departure}
\acrodef{itu}[ITU]{International Telecommunication Union}
\acrodef{dsm}[DSM]{differential spatial modulation}
\acrodef{smt}[SMT]{space modulation techniques}
\acrodef{dco}[DCO]{Asymmetrically clipped DC biased optical}
\acrodef{aco}[ACO]{Asymmetrically clipped optical}
\acrodef{ofdm}[OFDM]{orthogonal frequency division multiplexing}
\acrodef{papr}[PAPR]{Peak average power ratio}
\acrodef{mamimo}[MaMIMO]{massive multiple input multiple output}
\acrodef{ofdm}[OFDM]{orthogonal frequency division}
\acrodef{das}[DAS]{distributed antenna systems}
\acrodef{thz}[THz]{Terahertz}
\acrodef{aps}[APs]{access points}
\acrodef{ues}[UEs]{user equipment's}
\acrodef{6}[6G]{sixth generation}
\acrodef{uav}[UAV]{unmanned aerial vehicle }

\begin{document}

\title{FAS Enabled UAV for Energy-Efficient WPCNs }

\author{\IEEEauthorblockN{Nagla Abuzgaia\IEEEauthorrefmark{5}\IEEEauthorrefmark{7}, Abdelhamid Salem\IEEEauthorrefmark{5}\IEEEauthorrefmark{4},\textit{ Member, IEEE}, and Ahmed Elbarsha\IEEEauthorrefmark{5}}
\thanks{\IEEEauthorrefmark{5} Authors are with University of Benghazi, Electrical and Electronics Engineering Department, Faculty of Engineering, Benghazi, Libya, E-mails: \{{nagla.abuzgaia , abdelhamid.albaraesi \& ahmed.elbarsha\}@uob.edu.ly}}

\thanks{\IEEEauthorrefmark{7}Nagla Abuzgaia is also with Libyan Authority for Scientific Research, Tripoli, Libya, E-mail: {nagla.abuzgaia@aonsrt.ly}\\ \IEEEauthorrefmark{4} Abdelhamid Salem is also with Department of Electronic and Electrical Engineering, University College London, UK, E-mail: {a.salem@ucl.ac.uk}}}

\maketitle

\begin{abstract}
This letter presents an innovative scheme to enhance the communication rate and energy efficiency (EE) of Unmanned Aerial Vehicle (UAV) in wireless powered communication networks (WPCNs) by  deploying the emerging fluid antenna system (FAS) technology onto the UAV. Our proposed approach leverages the dynamic port switching capability of FAS, enabling the UAV to adaptively select the optimal antenna location that maximizes channel gain for both downlink wireless power transfer (WPT) and uplink wireless data transfer (WDT). We derive both exact analytical expression of the ergodic spectral rate,  and asymptotic expression at  high signal to noise ratio (SNR) regime under Nakagami-m correlated fading channels. The Mont-Carlo simulation results confirms the accuracy of the analytical expressions and demonstrate the substantial increase in energy efficiency of UAV with FAS compared to fixed antenna systems.

\end{abstract}


\IEEEpeerreviewmaketitle

\section{Introduction}

\IEEEPARstart{U}{nmanned} aerial vehicle (UAV) offers a paradigm shift in wireless communications, enabling on-demand, mobile connectivity with favourable line-of-sight (LoS) links. This capability is particularly transformative for wireless powered communication networks (WPCNs), where a UAV can serve as a mobile power beacon, wirelessly charging remote internet of things (IoT) devices that subsequently transmit their data back to the UAV \cite{xie2021uav}. Nevertheless, the efficacy of such systems is significantly hampered by the finite energy budget of the UAV and the efficiency challenges inherent to wireless power transfer. Fluid antenna systems (FAS) is an emerged technology that leverages the flexibility of port selection to harness diversity gain. Although the concept of FAS has recently gained traction; existing literature has not fully addressed its application in this context. Initial studies focused primarily on outage probability under Nakagami-m fading, often under simplified unit-distance assumptions \cite{tlebaldiyeva2022enhancing, vega2023simple}. While the ergodic capacity was explored in \cite{wong2020perf}, it was limited to Rayleigh fading, thus overlooking LoS effects. Other relevant works have either incorporated path loss using distinct methodologies such as copula theory for NOMA \cite{ghadi2024NOMA}, or analyzed FAS arrays too large for practical UAV deployment \cite{yang2024fast}. \\
In response to this, this letter introduces a novel framework that deploys FAS into UAV-enabled WPCNs to bolster energy efficiency and promote sustainable operation. The concept has been explored by deriving analytical expressions of ergodic spectral rate complemented by a high-SNR asymptotic analysis. The presented results validate our framework and demonstrate the tangible gains in energy efficiency and spectral rate afforded by the proposed system.

\section{system model}
\begin{figure}[h!]
 \includegraphics[width=8cm]{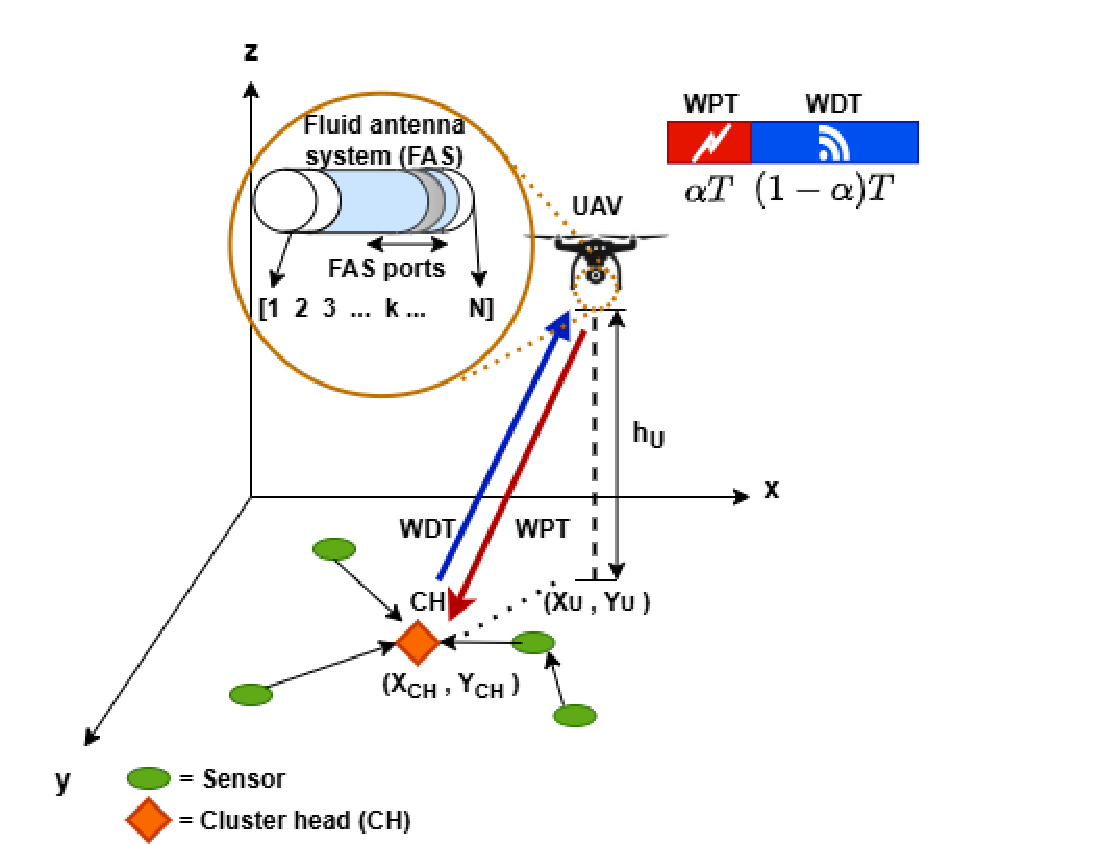}\vspace{-0.2cm}
 \caption{System model of a single FAS mounted UAV for WPCNs.}
 \label{fig:FA_UAV_WPT}
\end{figure}

In this study, to focus on the foundational performance metrics of the FAS mounted UAV in WPCNs, we consider an UAV hovering in a time slot $T$, representing an aerial base station (ABS) at the position coordinate $\mathbf{u} = (X_u, Y_u, h_u)^T \in \mathbb{R}^{3 \times 1}$ equipped with $N$-ports fluid antenna that transmits energy during $\alpha T$ to a designated ground node, which acts as a data aggregator for a cluster of sensors \cite{wei2022uav}, called cluster head (CH) and positioned at $\mathbf{CH}=(X_{CH},Y_{CH}, h)^T \in \mathbb{R}^{3 \times 1}$, in case the CH was placed on the ground then $h=0$. The distance between UAV and CH is $d=\sqrt{h_u^2 +(X_{CH}-X_u)^2+ (Y_{CH}-Y_u)^2}$. CH node employs harvest-then-transmit protocol, hence it transmits the collected data back to the UAV during the remaining of the time slot i.e. $(1-\alpha)T$, as demonstrated in Fig.\ref{fig:FA_UAV_WPT}.

\subsection{FAS-UAV channel model}
For analytical tractability, we assume the channel model as time division (TD). Due to the small size of the spacing between the N ports of the fluid antenna compared with the distance between the UAV and the CH we could approximate the $ k^{th}$ port pathloss of both the DL and UL as the same for all the ports, i.e. $L_k(d) \approx L(d) $, hence the total DL-WPT channel model and the UL-WDT channel model, respectively, are \cite{liu2021uav},
\begin{equation}
\{D_k, U_k\} = \sqrt{L(d)}\{h_{\bar{k}},g_{\tilde{k}}\} =\sqrt{\beta (d^{-\rho})}\{h_{\bar{k}},g_{\tilde{k}}\}  
\end{equation}
where $\beta$ is the channel power gain at the reference distance, $\rho$ is the path loss exponent, and $\{h_{\bar{k}} , g_{\tilde{k}}\} $ are the small-scale fading channels of the DL and  UL with $\{\bar{k},\tilde{k}\}$ as the activated port for each link, respectively. The fading channels are assumed to follow Nakagami-m distribution. Due to the small equipped dimension of the antenna the ports exhibit spatial correlation as follows \cite{mavrovoltsos2024fluid}:

\begin{equation}
\begin{split}
H_{kl} = \sqrt{1 - \mu^2_k} x_{kl} + \mu_k x_{0l} + j \left( \sqrt{1 - \mu^2_k} y_{kl} + \mu_k y_{0l} \right),\\
  l = \{1, \dots, m\},
 \end{split}
\end{equation}
where  $\{x_{kl}$, $ y_{kl}\}$ are independent Gaussian r.v.s with zero mean and 1/2 variance, $\{x_{ol}$, $ y_{ol}\}$ is the reference port, and $\mu_k$ is the correlation coefficient which follows \cite{wong2022closed}: 
\begin{equation}
\mu^2 = \frac{2}{N(N - 1)} \sum_{k=1}^{N-1} (N - k) J_0 \left( \frac{2\pi k W}{N - 1} \right), \quad \text{for } \mu_k = \mu \forall k.
\label{eq:mu_squared}
\end{equation}
where $J_0( )$ is the zero-order Bessel function of the first kind. Then the correlated $k^{th}$ port Nakagami-m fading channels envelops with normalized variance \cite{mavrovoltsos2024fluid}, 
\begin{equation}
|h_k| = \sqrt{\sum_{l=1}^{m} \frac{1}{m} |H_{kl}|^2}
\end{equation}
 
\subsection{FAS port selection strategy:} 
\subsubsection{Maximum Gain Selection (MGS)} 

The port selection strategy for both the WPT and the WDT is based on the FAS switching to the port with the maximum channel gain,\footnote{Considered simpler to implement and require less channel state information (CSI) feedback or processing by the UAV limited resources compared to maximizing the rate or maximizing SINR.} i.e: 
\begin{equation}
 |h_{\bar{k}}| = \arg \max_{k \in \mathcal{K}} \{|h_{k}|\} \quad \& \quad |g_{\tilde{k}}| = \arg \max_{k \in \mathcal{K}} \{|g_{k}|\}  
\end{equation}

To account for the correlation between FAS ports, the cumulative distribution function (CDF) of the selected channel gain is derived from the joint CDF of K dependent Nakagami-$m$ random variables presented in \cite{tlebaldiyeva2022enhancing}, and is expressed as:

\begin{equation}
\begin{split}
F_{|h_{\bar{k}}|}(|h_{FAS}|)=\frac{2m^{m}}{\Gamma(m)\sigma_{1}^{2m}}\int_{0}^{|h_{FAS}|}r_{1}^{2m-1}e^{-\frac{mr_{1}^{2}}{\sigma_{1}^{2}}}\\ \times\prod_{k=2}^{N}[1-Q_{m}(\sqrt{\frac{2m\mu_{k}^{2}r_{1}^{2}}{\sigma_{1}^{2}(1-\mu_{k}^{2})}},\sqrt{\frac{2m|h_{FAS}|^2}{\sigma_{k}^{2}(1-\mu_{k}^{2})}})]dr_{1}
\end{split}
\label{eqn:hfas}
\end{equation}
where $|h_{FAS}|$ represents the threshold for each individual  random variable $|h_k|$ , $m$ is the fading severity parameter, $\Gamma(m)$ is the Gamma function, $\sigma_{k}^{2}$ is the average channel power at port $k$, $\mu_{k}$ is the correlation coefficient of the k-th port with respect to the first port, and $Q_{m}(\cdot,\cdot)$ is the Marcum Q-function.

\subsubsection{Random Selection (RS)} 
This port selection strategy will be used as a benchmark to compare the performance of MGS method, where the activated port will be selected randomly without any criteria to present the performance of the fixed antenna systems\cite{mavrovoltsos2024fluid}.

\section{Received signal \& SNR }

The harvested energy in a duration of $\alpha T$ at the CH is \cite{liu2021uav}: 
\begin{equation}
E^{CH}_{\bar{k}} = \eta \,P_uL(d)|h_{\bar{k}}|^2 \alpha T
\end{equation}
where $\eta$ is the energy conversion efficiency, $P_u$ is the transmitted power of the UAV.\\
The CH uses the harvested energy for the UL data transfer to the UAV. The received data signal in a duration $(1-\alpha)T$ is:
\begin{equation}
{y}_{\tilde{k}} = \sqrt{\frac{\eta \,P_u \,  L^2(d)|h_{\bar{k}}|^2 \alpha }{(1 - \alpha)}} {g}_{\tilde{k}} {s} + n,
\end{equation}
where $\mathbf{s}$ is the normalized data symbol with zero mean and unit variance, and ${n}$ is the additive white Gaussian noise (AWGN) with zero mean and $(N_o)$ variance.\\
Thus the instantaneous SNR is: 

\begin{equation}
\gamma_r = \frac{ \eta  P_u  L^2(d)\alpha  \, |{h}_{\bar{k}}|^2   |{g}_{\tilde{k}}|^2 }{(1 - \alpha)N_o} 
\end{equation}

This leads to: 
\begin{equation}
\gamma_r = \nu|{h}_{\bar{k}}|^4  \quad \text{where} \quad \nu = \frac{\eta  P_u  L^2(d)\alpha}{(1 - \alpha)N_o} 
 \label{eqn:snr} 
\end{equation}

\section{performance analysis}
\subsection{Ergodic Spectral Rate}

\textit{Theorem1:} The exact integral form of the ergodic spectral rate, presented in (\ref{eqn:Rerg}) at the bottom of next page.
\begin{figure*}[b]
\rule{\linewidth}{0.8pt}
\begin{equation}
\begin{split}
R_{\text{ergodic}} &= \frac{(1 - \alpha)}{\ln(2)}
\int_{0}^{\infty} \frac{1}{1+\gamma_0} \left(1-  \frac{2m^{m}}{\Gamma(m)\sigma_{1}^{2m}}\int_{0}^{\left(\frac{\gamma_0}{\nu}\right)^{1/4}}r_{1}^{2m-1}e^{-\frac{mr_{1}^{2}}{\sigma_{1}^{2}}}\right. \\ 
&\quad \left. \times\prod_{k=2}^{N} \left[1-Q_{m} \left(\sqrt{\frac{2m\mu_{k}^{2}r_{1}^{2}}{\sigma_{1}^{2}(1-\mu_{k}^{2})}},\sqrt{\frac{2m \sqrt{\frac{\gamma_0}{\nu}}}{\sigma_{k}^{2}(1-\mu_{k}^{2})}}\right)\right]dr_{1} \right) d\gamma_0
\end{split}
\label{eqn:Rerg}
\end{equation}
\end{figure*}

\textit{Proof:} 
The ergodic spectral rate is calculated by  
 \begin{equation}
R_{\text{ergodic}} =\frac{(1 - \alpha)}{\ln(2)} \int_{0}^{\infty} \frac{1}{(1 + \gamma_o)} (1 - F_{\gamma_r}(\gamma_o)) d\gamma_o
\label{eqn:rergodic}
\end{equation} 
  
Referring to the equations (\ref{eqn:hfas}) and (\ref{eqn:snr}), for the CDF of the $\bar{k}^{th} $ port FA-UAV channel  $ F_{|h_{\bar{k}}|}$ and the instantaneous SNR $\gamma_r$, respectively, and using variable transformation as follows: 

\begin{equation}
F_{\gamma_r}(\gamma_0) = P(\gamma_r \le \gamma_0) = P(\nu |h_{\bar{k}}|^4 \le \gamma_0).
\end{equation}

\begin{equation}
  \quad P\left(|h_{\bar{k}}|^4 \le \frac{\gamma_0}{\nu}\right) = P\left(|h_{\bar{k}}| \le \left(\frac{\gamma_0}{\nu}\right)^{1/4}\right),\gamma_0 \ge 0 
\end{equation}

\begin{equation}
F_{\gamma_r}(\gamma_0) = F_{|h_{\bar{k}}|}\left(\left(\frac{\gamma_0}{\nu}\right)^{1/4}\right).
\label{eqn:Fgamma}
\end{equation}

Using $F_{|h_{\bar{k}}|}$ formulas in (\ref{eqn:hfas}) and  (\ref{eqn:Fgamma}) into (\ref{eqn:rergodic}), we get (\ref{eqn:Rerg}).

\subsection{Ergodic spectral rate at high SNR}

 \textit{Theorem2:} The asymptotic ergodic spectral rate at high SNR is expressed in (\ref{eqn:R_high_fin}) below.
 \begin{figure*}[b]
 \begin{equation}
R^{high}_{ergodic} \approx \frac{1-\alpha}{\ln(2)} \left[ \ln\left(\frac{\eta P_u L^2(d)\alpha}{(1-\alpha)N_o}\right)+ 2\psi(mN) - \frac{2}{mN} \ln\left(\Gamma(mN) \cdot a_0 \cdot mN\right) \right]
\label{eqn:R_high_fin}
\end{equation}

\end{figure*}
 
 \textit{Proof:} For the asymptotic high SNR, $\gamma_r \to \infty$,

 \begin{equation}
R^{high}_{\text{ergodic}} \approx \frac{(1 - \alpha)}{ln(2)}\mathbb{E} [ ln(\gamma_r)]
\label{eqn:R_high}
\end{equation}

By using (\ref{eqn:Fgamma}) and the approximation of $F_{|h_{\bar{k}}|^2} $ found in \cite[Eq. (8)]{vega2023simple}:       
\begin{equation}
F_{\gamma_r}(\gamma_o) \approx  \left(\frac{\gamma(s, \frac{\sqrt{\gamma_0}}{\beta \sqrt{\nu}})}{\Gamma(s)}\right)
\label{eqn:F_gamma}
\end{equation}
 
\begin{align*}
\text{where} \quad  s &= mN, \quad \beta = \left( \frac{1} { \Gamma(s) a_o s}\right)^{1/s},   \\
a_0 &= \frac{m^{m-1}}{\Gamma(m)\sigma_{1}^{2m} m!^{N-1}} \prod_{k=2}^{N} \left( \frac{m}{ \sigma_{k}^2(1 - \mu_k^2)} \right)^m 
\end{align*}
$\gamma(.)$ and $\Gamma(.)$ are the lower incomplete gamma function and gamma function, respectively. Both represent the CDF of $\sqrt{\gamma_r}$ that follows Gamma distribution with the shape and scale parameters defined as $(k=s,\theta=\beta\sqrt{\nu})$. Thus, 
\begin{equation}
\mathbb{E}[ln(\gamma_r)]=2[\psi(k)+ ln(\theta)]=2\psi(s)+ln(\nu)-\frac{2}{s}ln(\Gamma(s)a_os)
\end{equation}
where $\psi(.)$ is the digamma function. Substitution in (\ref{eqn:R_high}) ends the proof.\\
 \textbf{Remark:} \textit{The derived expression deconstructs the system's performance into three distinct analytical components. The first term quantifies the system power gain, governed by the UAV's transmit power and the link's path loss. The second term represents the gross diversity gain, which is a function of the total diversity order ($s=mN$), capturing the combined benefits of the Nakagami fading parameter (m) and the number of FAS ports(N). The final term introduces a fading and correlation penalty, a sophisticated adjustment factor dependent on the structural parameter ($a_o$), which encapsulates the limiting effects of inter-port channel correlation and the intrinsic fading severity.}
\subsection{Energy Efficiency}
The energy efficiency ($\zeta$) of FAS-UAV in WPCN is defined as the ergodic rate at the optimal time splitting ratio $\alpha$ divided by the total UAV consumption power\cite{xu2023energy}
\begin{equation}
\zeta = R_{ergodic}/P_{tot}
\end{equation}
where ${P_{tot} = P_c + P_u+P_h }$, $P_c$ is the constant communication circuit power, $P_u$ is the UAV power amplifier consumed transmitted power, and $P_h= P_o +P_i$ is the power consumed by the rotary-wing UAV to stay in the hovering status\cite{zeng2019energy}, given $P_o$ is the blade profile power and $P_i$ is the induced power.

\section{Simulation Results}
\begin{figure}[h!]
 \centering
 \includegraphics[width=8cm]{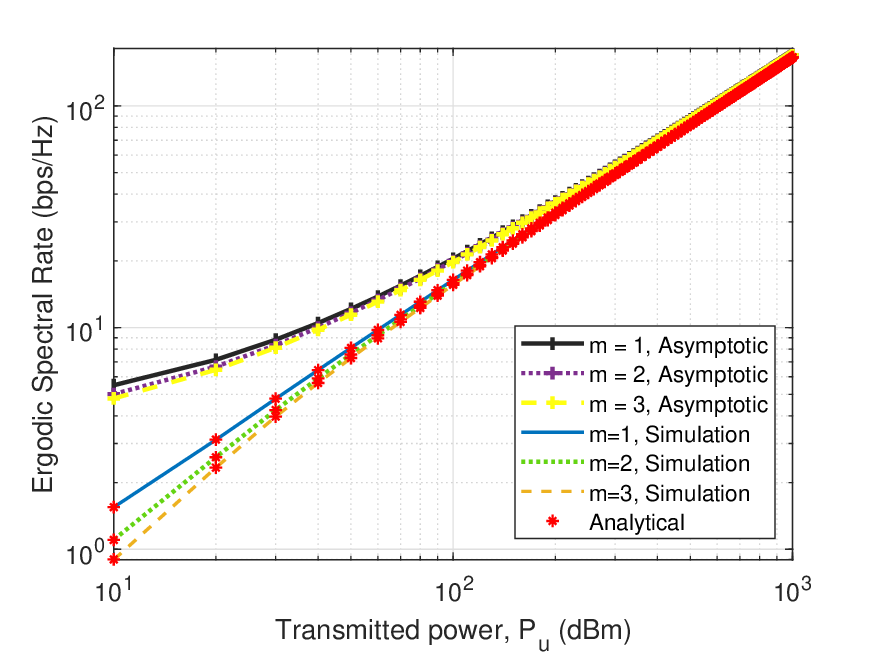}\vspace{-0.2cm}
 \caption{The spectral efficiency vs. transmitted power($P_u$) for ($\alpha=0.5$, $K=200$ports,$W=2$, and $d=25m$ ) with Nakagami fading $(m=1,2,3)$.} 
\label{fig:high_snr}
\end{figure}

\begin{figure}[h!]
 \centering
 \includegraphics[width=8cm]{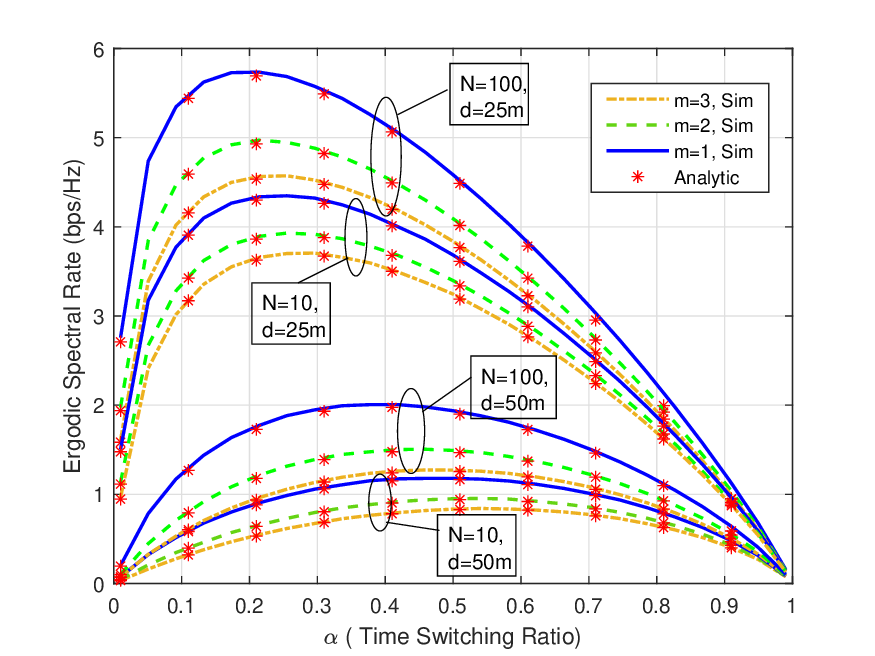}\vspace{-0.2cm}
 \caption{Ergodic spectral rate vs. the time switching ratio ($\alpha$) for distances $d=(25,50)m$, FAS ports $N =(10,100)$ and Nakagami-m fading $(m=1,2)$.}
\label{fig:erg_C_vs_TS}
\end{figure}

\begin{figure}[th!]
 \centering
 \includegraphics[width=8cm]{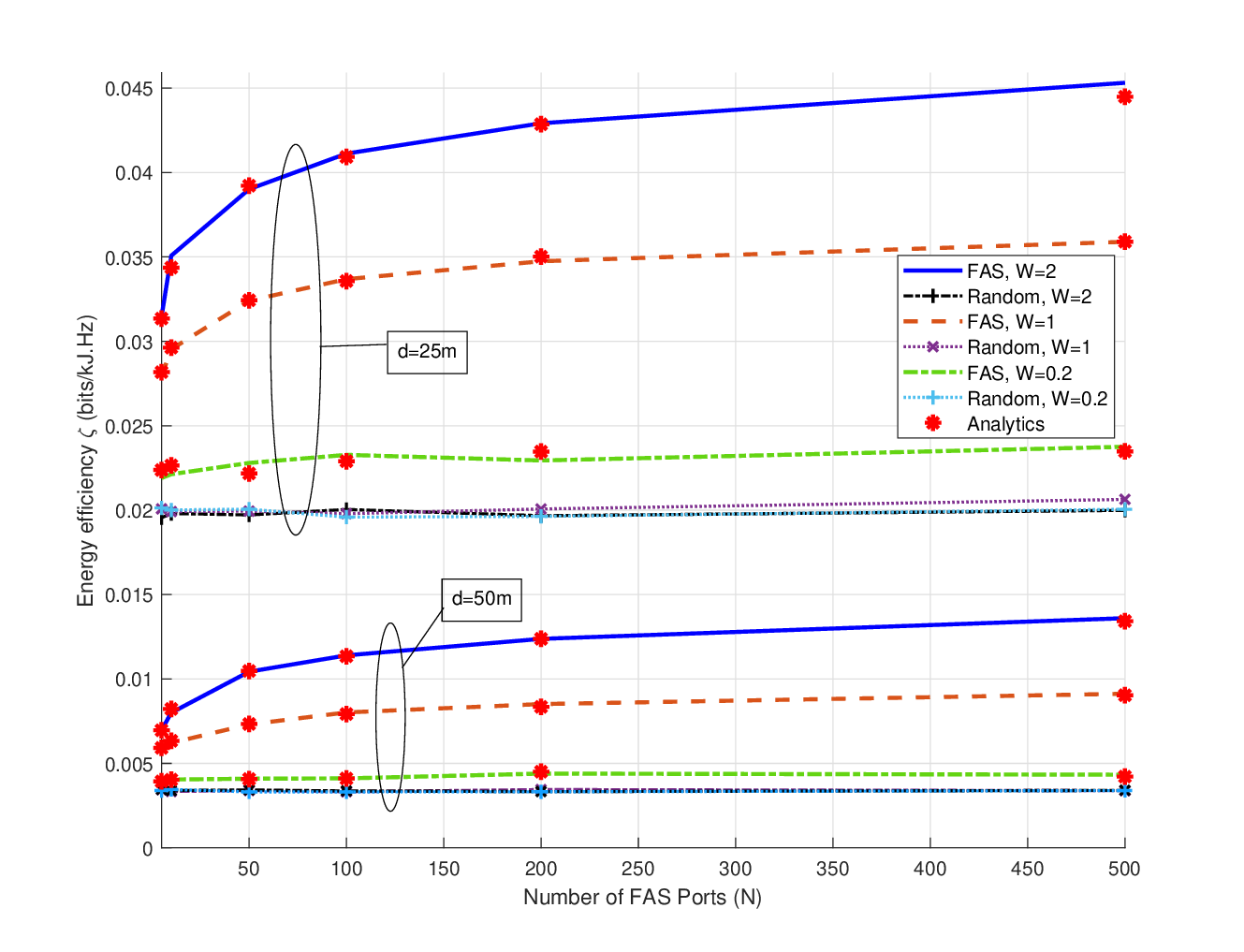}\vspace{-0.2cm}
 \caption{The energy efficiency at the optimal time switching ratio ($\alpha$) for distances $d=(25,50)m$, FAS sizes $W=(0.2,1,2)\lambda$,  and  Nakagami fading $(m=2)$.} 
\label{fig:erg_C_vs_k}
\end{figure}

The performance of a FAS-mounted UAV in WPCN is evaluated through Monte Carlo simulations with a FAS width of $W=2\lambda$. The UAV transmission power is $P_u=1 W$ and the EH efficiency $\eta=0.8$.  The wireless channel is modelled as a Nakagami-$m$ fading channel with a path loss exponent of $\rho=2.7$ and  AWGN power of $N_o=10^{-9}W$.\\ 
 Fig.\ref{fig:high_snr} plots the ergodic spectral rate vs. the UAV transmitted power, validating the analytical framework. It showed a perfect match between the derived analytical expressions for ergodic spectral rate and the simulation results across a wide range of transmitted power levels, $P_u$. Furthermore, it confirms the accuracy of our high SNR analysis, as the asymptotic curves are shown to converge with the analytical results in the high-power regime ($P_u > 10^2 dBm$).\\ 
Fig.\ref{fig:erg_C_vs_TS} illustrates the relationship between the ergodic spectral rate and the time-switching ratio, $\alpha$, highlighting the impact of key system parameters. Increasing the number of FAS ports from $K=10$ to $K=100$ yields a substantial increase in the peak spectral rate, demonstrating the diversity gain inherent to FAS. 
Similarly, reducing the UAV-to-CH distance from $d=50m$ to $d=25m$ provides a significant vertical uplift in performance, a direct result of the enhanced system power gain. Crucially, both improvements lead to a more energy-efficient system, evidenced by the shift of the optimal $\alpha$ to the left. This indicates that a shorter relative energy harvesting time is required to achieve a significantly higher spectral efficiency. A consistent and notable observation across both figures is the ergodic spectral rate decreases as the Nakagami-$m$ fading parameter increases. The phenomenon can be explained by the channel's coefficient of variation (CV). A lower value of $m$ (more severe fading) corresponds to a larger CV, which increases the probability of the FAS selecting an exceptionally strong channel realization. These infrequent but large channel gains dominate the long-term average rate. Conversely, a higher $m$ leads to a more deterministic channel with less variation, reducing the peaks available for the FAS to exploit.\\ 
Finally, Fig.\ref{fig:erg_C_vs_k} demonstrates the energy efficiency gain of the intelligent FAS selection over a random port selection strategy under different distances and FAS sizes, due to the increase of spectral rate a less UAV hovering time is required for the same task. Decreasing the FAS size $W$ clearly demonstrate lower energy efficiency due to the correlation effect, which further elucidated by the penalty term in the high SNR equation. The energy efficiency grows logarithmically with the number of FAS ports $(N)$, exhibiting diminishing returns as N becomes large. This figure also powerfully illustrates the impact of path loss, where decreasing the distance from $50m$ to $25m$ results in a multi-fold increase in performance, reinforcing the importance of the system power gain in FAS-UAV enabled networks.   
  
\section{Conclusion}
This study analysed the potential of the FAS for optimizing the ergodic spectral rate and energy  efficiency of the UAV in WPCNs. It provided a crucial first step in understanding the benefits and challenges of employing FAS in this context, an exact analytical equation for the ergodic spectral rate was provided. Along an asymptotic high SNR equation that was able to unleash the effect of the FAS parameters, path-loss and severity of fading on the system performance. A single FAS substantially increased the energy efficiency compared with fixed antenna systems, proving as a promising solution for the WPCNs as well for the limited battery and weight barrier of the UAV.
  
\bibliographystyle{IEEEtran}
\bibliography{FA_WPT_UAV_fin} 

\begin{thebibliography}{10}
\providecommand{\url}[1]{#1}
\csname url@samestyle\endcsname
\providecommand{\newblock}{\relax}
\providecommand{\bibinfo}[2]{#2}
\providecommand{\BIBentrySTDinterwordspacing}{\spaceskip=0pt\relax}
\providecommand{\BIBentryALTinterwordstretchfactor}{4}
\providecommand{\BIBentryALTinterwordspacing}{\spaceskip=\fontdimen2\font plus
\BIBentryALTinterwordstretchfactor\fontdimen3\font minus
  \fontdimen4\font\relax}
\providecommand{\BIBforeignlanguage}[2]{{%
\expandafter\ifx\csname l@#1\endcsname\relax
\typeout{** WARNING: IEEEtran.bst: No hyphenation pattern has been}%
\typeout{** loaded for the language `#1'. Using the pattern for}%
\typeout{** the default language instead.}%
\else
\language=\csname l@#1\endcsname
\fi
#2}}
\providecommand{\BIBdecl}{\relax}
\BIBdecl

\bibitem{xie2021uav}
L.~Xie, X.~Cao, J.~Xu, and R.~Zhang, ``Uav-enabled wireless power transfer: A
  tutorial overview,'' \emph{IEEE Transactions on Green Communications and
  Networking}, vol.~5, no.~4, pp. 2042--2064, 2021.

\bibitem{tlebaldiyeva2022enhancing}
L.~Tlebaldiyeva, G.~Nauryzbayev, S.~Arzykulov, A.~Eltawil, and T.~Tsiftsis,
  ``Enhancing qos through fluid antenna systems over correlated nakagami-m
  fading channels,'' in \emph{IEEE Wireless Communications and Networking
  Conference (WCNC)}, 2022, pp. 78--83.

\bibitem{vega2023simple}
J.~D. Vega-S{\'a}nchez, L.~Urquiza-Aguiar, M.~C.~P. Paredes, and D.~P.~M.
  Osorio, ``A simple method for the performance analysis of fluid antenna
  systems under correlated nakagami-m fading,'' \emph{IEEE Wireless
  Communications Letters}, vol.~13, no.~2, pp. 377--381, 2023.

\bibitem{wong2020perf}
K.~K. Wong, A.~Shojaeifard, K.-F. Tong, and Y.~Zhang, ``Performance limits of
  fluid antenna systems,'' \emph{IEEE Communications Letters}, vol.~24, no.~11,
  pp. 2469--2472, 2020.

\bibitem{ghadi2024NOMA}
F.~R. Ghadi, M.~Kaveh, K.-K. Wong, R.~J{\"a}ntti, and Z.~Yan, ``On performance
  of fas-aided wireless powered noma communication systems,'' in \emph{20th
  International Conference on (WiMob)}.\hskip 1em plus 0.5em minus 0.4em\relax
  IEEE, 2024, pp. 496--501.

\bibitem{yang2024fast}
H.~Yang, X.~Lin, K.-K. Wong, and Y.~Zhao, ``Fast fluid antenna multiple access
  with path loss consideration and different antenna architecture,'' in
  \emph{Proc. 22nd International Conference on Trust, Security and Privacy in
  Computing and Communications}.\hskip 1em plus 0.5em minus 0.4em\relax IEEE,
  2024, pp. 2386--2393.

\bibitem{wei2022uav}
Z.~Wei, M.~Zhu, N.~Zhang, L.~Wang, Y.~Zou, Z.~Meng, H.~Wu, and Z.~Feng,
  ``Uav-assisted data collection for internet of things: A survey,'' \emph{IEEE
  Internet of Things Journal}, vol.~9, no.~17, pp. 15\,460--15\,483, 2022.

\bibitem{liu2021uav}
Y.~Liu, K.~Xiong, Y.~Lu, Q.~Ni, P.~Fan, and K.~B. Letaief, ``Uav-aided wireless
  power transfer and data collection in rician fading,'' \emph{IEEE Journal on
  Selected Areas in Communications}, vol.~39, no.~10, pp. 3097--3113, 2021.

\bibitem{mavrovoltsos2024fluid}
T.~Mavrovoltsos, E.~Demarchou, C.~Psomas, and I.~Krikidis, ``Fluid antenna
  systems for terahertz wireless power transfer,'' in \emph{GLOBECOM
  2024}.\hskip 1em plus 0.5em minus 0.4em\relax IEEE, 2024, pp. 3176--3181.

\bibitem{wong2022closed}
K.~Wong, K.~Tong, Y.~Chen, and Y.~Zhang, ``Closed-form expressions for spatial
  correlation parameters for performance analysis of fluid antenna systems,''
  \emph{Electronics Letters}, vol.~58, no.~11, pp. 454--457, 2022.

\bibitem{xu2023energy}
Y.~Xu, Y.~Chen, Y.~Hou, K.-K. Wong, Q.~Cui, and X.~Tao, ``Energy efficiency
  maximization under delay-outage probability constraints using fluid antenna
  systems,'' in \emph{IEEE Statistical Signal Processing Workshop (SSP)}.\hskip
  1em plus 0.5em minus 0.4em\relax IEEE, 2023, pp. 105--109.

\bibitem{zeng2019energy}
Y.~Zeng, J.~Xu, and R.~Zhang, ``Energy minimization for wireless communication
  with rotary-wing uav,'' \emph{IEEE transactions on wireless communications},
  vol.~18, no.~4, pp. 2329--2345, 2019.

\end{thebibliography}
\end{document}